\documentclass[runningheads,a4paper]{llncs}

\usepackage{amssymb}
\usepackage{bm,bbm}
\usepackage{graphicx,subfigure}
\usepackage{booktabs}
\usepackage{multirow}
\usepackage[ugly]{units}
\usepackage{microtype}

\usepackage{url}
\urldef{\mails}\path|ljmtorres, tamersgc, acfrery}@gmail.com|    
\newcommand{\keywords}[1]{\par\addvspace\baselineskip
\noindent\keywordname\enspace\ignorespaces#1}

\begin{document}
\title{Speckle Reduction using Stochastic Distances}

\titlerunning{Speckle Reduction using Stochastic Distances}

\author{Leonardo Torres\and Tamer Cavalcante \and Alejandro C.\ Frery \thanks{The authors are grateful to CNPq and Fapeal for supporting this research}}

\authorrunning{Torres, Cavalcante \& Frery}

\institute{
Universidade Federal de Alagoas -- UFAL\\
Laborat\'orio de Computa\c c\~ao Cient\'ifica e An\'alise Num\'erca -- LaCCAN\\
57072-970, Macei\'o, AL -- Brazil
}

\toctitle{Speckle Reduction using Stochastic Distances}
\tocauthor{Torres, Cavalcante \& Frery}
\maketitle

\begin{abstract}
This paper presents a new approach for filter design based on stochastic distances and tests between distributions.
A window is defined around each pixel, samples are compared and only those which pass a goodness-of-fit test are used to compute the filtered value.
The technique is applied to intensity Synthetic Aperture Radar (SAR) data, using the Gamma model with varying number of looks allowing, thus, changes in heterogeneity.
Modified Nagao-Matsuyama windows are used to define the samples.
The proposal is compared with the Lee's filter which is considered a standard, using a protocol based on simulation.
Among the criteria used to quantify the quality of filters, we employ the equivalent number of looks (related to the signal-to-noise ratio), line contrast, and edge preservation.
Moreover, we also assessed the filters by the Universal Image Quality Index and the Pearson's correlation between edges.
\end{abstract}

\section{Introduction}
\label{sec:intro}

SAR plays an important role in Remote Sensing since they provide complementary information to that provided by optical sensors.
SAR data are subjected to speckle noise, which is also present in laser, ultrasound-B, and sonar imagery~\cite{Goodman1976}. 
This noise degrades the SAR information content and makes image in\-ter\-pre\-ta\-tion clas\-si\-fi\-ca\-tion difficult~\cite{Lee1986}.

Statistical analysis is essential for dealing with speckle.
In addition, statistical modeling provides support for the development of algorithms for in\-ter\-preting the data efficiently, and for the simulation of plausible images.
Different statistical distributions are proposed in the literature to describe speckled data.
We used the multiplicative model in intensity format for ho\-mo\-ge\-ne\-ous areas, ergo the Gamma distribution was employed to describe the data~\cite{Gao2010}.

This work presents new filters based on stochastic distances and tests between distributions, as presented in Nascimento et al.~\cite{Nascimento2010}.
The filters are compared to Lee's filter using a protocol proposed by Moschetti et al.~\cite{Moschetti2006} using Monte Carlo simulation.
The criteria used to evaluate this filters are the equivalent number of looks, line contrast, edge preserving, the $Q$ index~\cite{UIQIndex} and Pearson's correlation between edges.

The paper is organized as follows:
In Section~\ref{sec:model} we summarise the model for speckle data.
Section~\ref{sec:distances} describe the new filters.
Section~\ref{sec:assessment} presents the measures for assessing the quality of filtered images,
with conclusions drawn in Section~\ref{sec:conclu}.

\section{The Multiplicative Model}\label{sec:model}

According to Goodman~\cite{Goodman1976}, the multiplicative model can be used to describe SAR data.
This model asserts that the intensity observed in each pixel is the outcome of the random variable $Z\colon\Omega \rightarrow \mathbbm{R}_+$ which is the product of two independent random variables: 
$X\colon\Omega \rightarrow \mathbbm{R}_+$, that characterizes the backscatter; and
$Y\colon\Omega \rightarrow \mathbbm{R}_+$, which models the speckle noise.
The distribution of the observed intensity $Z=XY$ is completely specified by the distributions of $X$ and $Y$.

This proposal deals with homogeneous regions in intensity images, so a con\-stant $X\thicksim\lambda>0$ defines the backscatter, and the speckle noise is described by a Gamma distribution $Y\thicksim \Gamma(L,L)$ with unitary mean $\mathbbm{E}(Y)=1$, 
where $L\geq1$ is number of looks.
Thus, it follows that $Z\thicksim \Gamma(L,{L}/{\lambda})$ with density
\begin{equation}
f_Z(z;L,\lambda) = \frac{L^L}{\lambda^L\Gamma(L)} z^{L-1} \exp\Big\{ \frac{-Lz}{\lambda} \Big\},
\label{eq:densGamma}
\end{equation}
where $\Gamma$ is the gamma function, $z,\lambda > 0$ and $L\geq1$.

\section{Stochastic Distances Filter}\label{sec:distances}

\begin{figure}[hbt]
\centering
  \subfigure[$5\times5$ neighborhood]{\label{fig:NM_w5}
  \includegraphics[width=.48\linewidth]{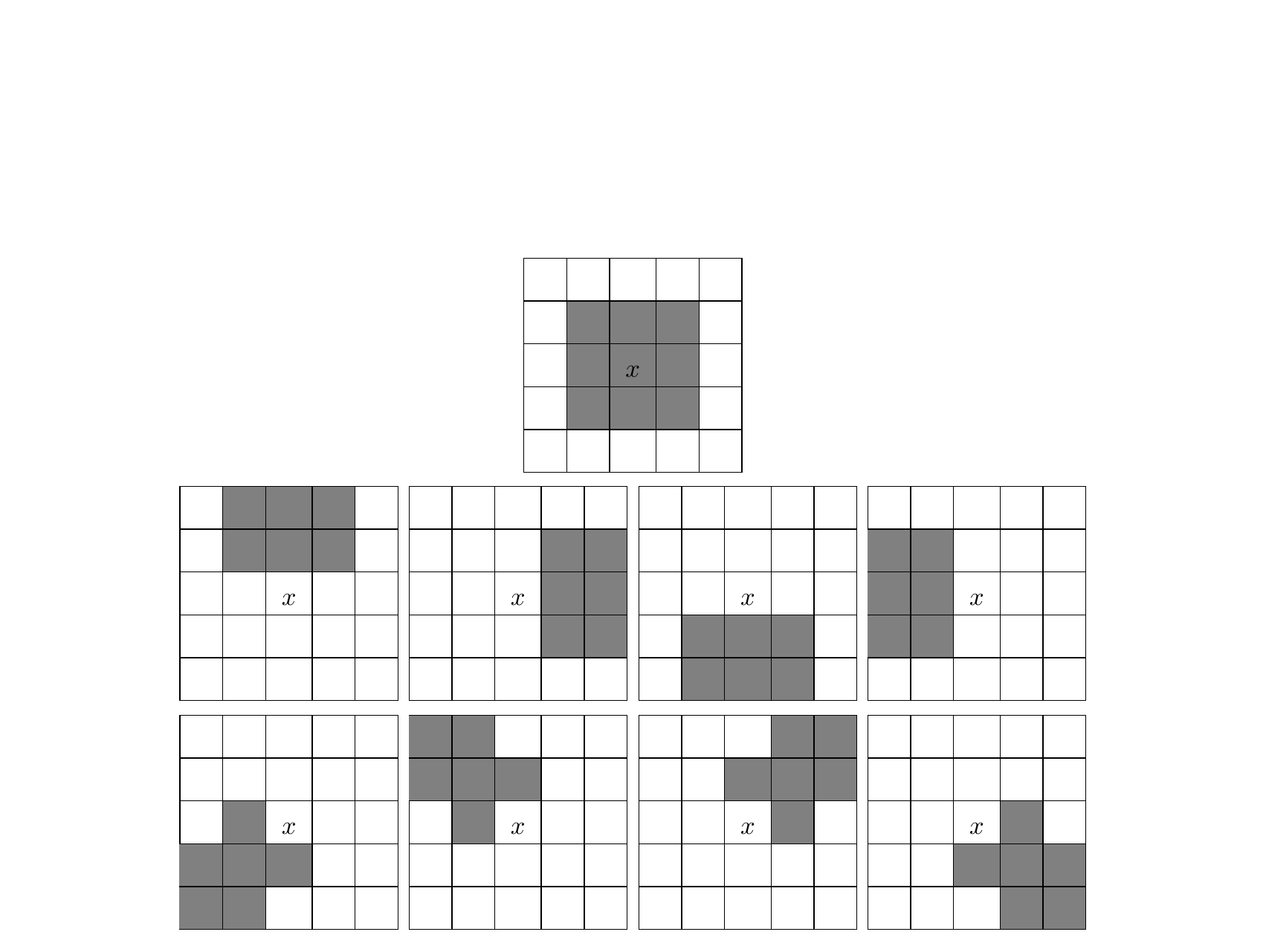}}
  \subfigure[$7\times7$ neighborhood]{\label{fig:NM_w7}
  \includegraphics[width=.48\linewidth]{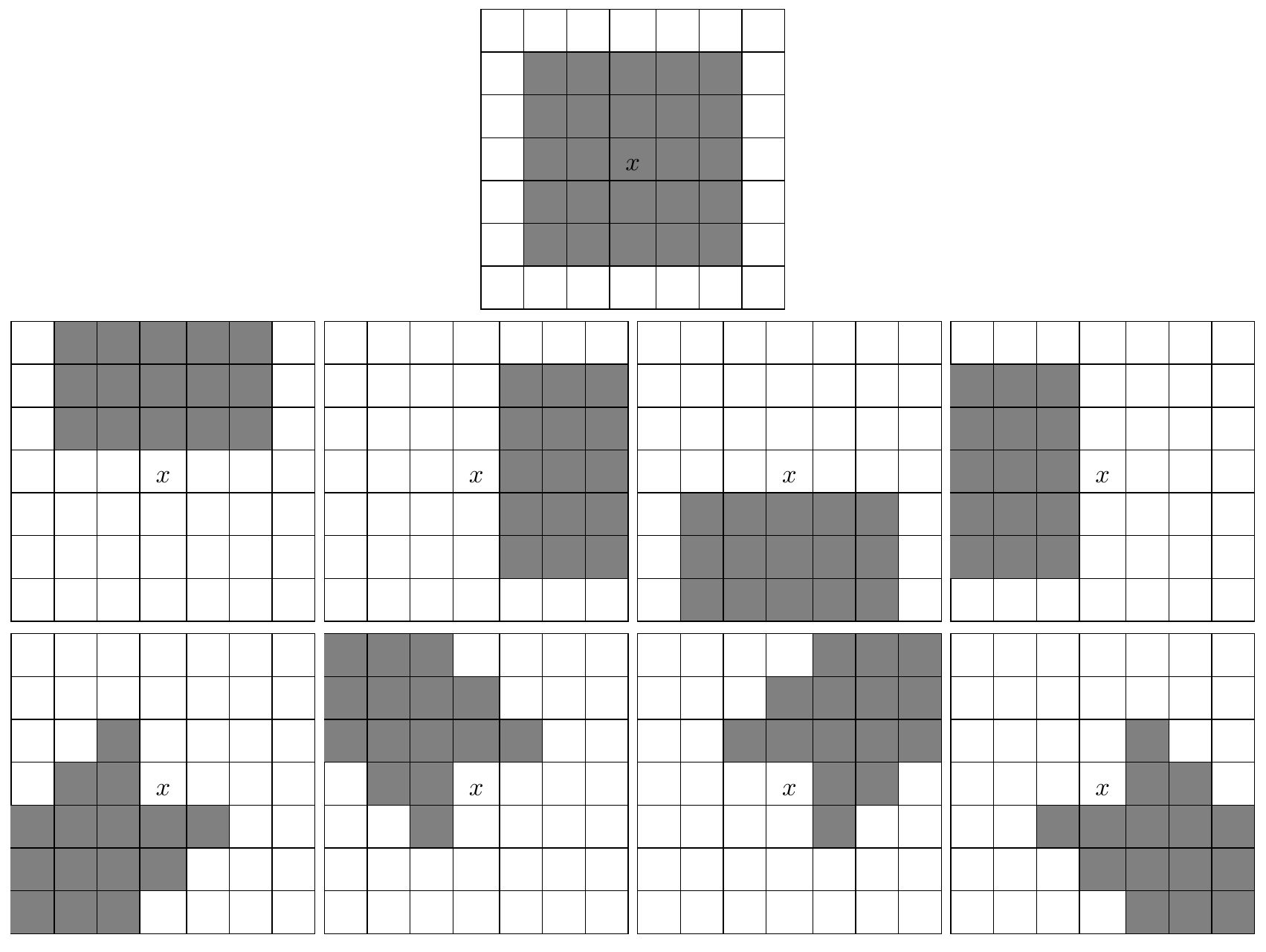}}
\caption{Nagao-Matsuyama neighbourhoods.}
\label{fig:NagaoMatsuyama}
\end{figure}

The proposed filter is based on stochastic distances and tests between dis\-tri\-bu\-tions~\cite{Nascimento2010}, obtained from the class of ($h,\phi$)-divergences.
It employs neigh\-bor\-hoods as defined by Nagao and Matsuyama~\cite{NagaoMatsuyama}, presented in Figure~\ref{fig:NM_w5}, and extended versions as shown in Figure~\ref{fig:NM_w7}.

Each filtered pixel has a $5\times5$ neighborhood (see Figure~\ref{fig:NM_w5}) or a $7\times7$ neighborhood (see Figure~\ref{fig:NM_w7}), within which nine disjoint areas are defined.
Denote $\bm{\widehat{\theta}_1}$ the estimated parameter in the central $3\times3$ or $5\times5$ neigh\-bor\-hood, respectively, and $\big(\bm{\widehat{\theta}}_2,\ldots,\bm{\widehat{\theta}}_{9}\big)$ the estimated parameters in the eight re\-main\-ing areas.
To account for possible departures from the homogeneous model, we estimate $\bm{\widehat{\theta}}_i=(L_i,\lambda_i)$ by maximum likelihood; reduced equivalent number of looks are related to heterogeneous areas~\cite{EstimationEquivalentNumberLooksSAR}.

Based on a random sample of size $n$, $\bm{z}=(z_1,z_2,\dots,z_n)$, 
the likelihood function related to the $\Gamma(L,L/\lambda)$ distribution is given by
\begin{equation}
\mathcal{L}(L,\lambda;\bm{z})=\bigg( \frac{L^L}{\lambda^L\Gamma(L)} \bigg)^n \prod_{j=1}^{n} z_j^{L-1} \exp\Big\{ \frac{-Lz_j}{\lambda} \Big\}.
\end{equation}
Thus, maximum likelihood estimators for $(L,\lambda)$, namely, $(\widehat{L},\widehat{\lambda})$, are the solution of the following system of non-linear equations:
$$
\left\{
\begin{array}{r}
\ln\widehat{L} - \psi^0(\widehat{L}) - \ln\big( \frac{1}{n}\sum_{j=1}^{n}z_j \big) + \frac{1}{n}\sum_{j=1}^{n} \ln z_j = 0, \\
-\frac{n\widehat{L}}{\widehat{\lambda}}+\frac{\widehat{L}}{\widehat{\lambda}^2}\sum_{j=1}^{n}z_j = 0,
\end{array}
\right.
$$
where $\psi^0$ is the digamma function.

The proposal is based on the use of stochastic distances on small areas within the filtering window.
Consider $Z_1$ and $Z_i$ random variables defined on the same probability space, whose distributions are characterized by the den\-si\-ties $f_{Z_1}(z_1;\bm{\theta}_1)$ and $f_{Z_i}(z_i;\bm{\theta}_i)$, res\-pec\-ti\-ve\-ly, where $\bm{\theta}_1$ and $\bm{\theta}_i$ are parameters.
Assuming that the densities have the same support $I \subset \mathbbm{R}$, the $h$-$\phi$ divergence between $f_{Z_1}$ and $f_{Z_i}$ is given by
\begin{equation}
D_{\phi}^{h}(Z_1,Z_i) = h \Big( \int_{x\in I}\;\phi \Big( \frac{f_{Z_1}(x;\bm{\theta}_1)}{f_{Z_i}(x;\bm{\theta}_i)} \Big) \;f_{Z_i}(x;\bm{\theta}_i)\;\mathrm{d}x \Big),
\end{equation}
where $h\colon(0,\infty)\rightarrow[0,\infty)$ is a strictly increasing function with $h(0)=0$ and $h'(x)>0$ for every $x \in \mathbbm{R}$, and $\phi\colon (0,\infty)\rightarrow[0,\infty)$ is a convex function~\cite{Salicru1994}.
Choices of functions $h$ and $\phi$ result in several divergences.

Divergences sometimes are not dis\-tan\-ces because they are not symmetric.
A simple solution, described in~\cite{Nascimento2010}, is to define a new measure $d_{\phi}^{h}$ given by
$
d_{\phi}^{h}(Z_1,Z_i) = \big({D_{\phi}^{h}(Z_1,Z_i)+D_{\phi}^{h}(Z_i,Z_1)}\big)/ 2.
$
Distances, in turn, can be conveniently scaled to present good statistical prop\-er\-ties that make them statistical hypothesis tests~\cite{Nascimento2010}:
\begin{equation}
S_{\phi}^{h}(\bm{\widehat{\theta}}_1,\bm{\widehat{\theta}}_i)=\frac{2mnk}{m+n}\;d^{h}_{\phi}(\bm{\widehat{\theta}}_1,\bm{\widehat{\theta}}_i),
\end{equation}
where $\bm{\widehat{\theta}}_1$ and $\bm{\widehat{\theta}}_i$ are maximum likelihood estimators based on samples size $m$ and $n$, respectively, and $k=(h'(0)\phi'') ^{-1}$.
When $\bm\theta_1=\bm\theta_i$, under mild conditions $S_{\phi}^{h}(\bm{\widehat{\theta}}_1,\bm{\widehat{\theta}}_i)$ is asymptotically $\chi^2_M$  distributed, being $M$ the dimension of $\bm{\theta}_1$.
Ob\-serv\-ing $S_{\phi}^{h}(\bm{\widehat{\theta}}_1,\bm{\widehat{\theta}}_i)=s$, the null hypothesis $\bm{\theta}_1=\bm{\theta}_i$ can be rejected at level $\eta$ if $\Pr( \chi^2_{M}>s)\leq \eta$.
Details can be seen in the work by Salicr\'u et al.~\cite{Salicru1994}.

Since we are using the same sample for eight tests, we modified the value of $\eta$ by a Bonferroni-like correction, namely, \u{S}id\'ak correction, that is given by $\eta = 1 - (1 - \alpha)^{1/t}$, where $t$ is the number of tests and, $\alpha$ the level of significance for the whole series of tests.

Nascimento et al.~\cite{Nascimento2010} derived several distances for the $\mathcal G^0$ model, which includes the one presented in Equation~(\ref{eq:densGamma}).
We opted for the latter, due to the numerical complexity of the former; the lack of flexibility is alleviated by allowing the number of looks to vary locally.
The statistical tests used in this paper are then:
\begin{description}
 \item[Hellinger test:] $S_{H} = \frac{8mn}{m+n}\bigg(1-\frac{2^{\widehat{L}}(\widehat{\lambda}_1\widehat{\lambda}_i)^{\widehat{L}/2}}{(\widehat{\lambda}_1+\widehat{\lambda}_i)^{\widehat{L}}}\bigg).$

 \item[Kulback-Leibler test:] $S_{KL} = \frac{2mn}{m+n}\;\widehat{L}\bigg(\frac{\widehat{\lambda}_1^2+\widehat{\lambda}_i^2}{2\widehat{\lambda}_1 \widehat{\lambda}_i}-1\bigg).$

 \item[R\'enyi test of order $\beta$:] $S_{R}^{\beta} = \frac{2mn}{m+n} \frac{\widehat{L}}{2\beta(\beta-1)}\;
 \log\frac{\widehat{\lambda}_1\widehat{\lambda}_i}{\big( \beta\widehat{\lambda}_i+(1-\beta)\widehat{\lambda}_1\big)\;\big(\beta\widehat{\lambda}_1+(1-\beta)\widehat{\lambda}_i \big)},$ in which $0 < \beta < 1$.
\end{description}

Although these are all different tests, in practice they led to exactly the same decisions in all situations here considered.
We, therefore, chose to work only with the test based on the Hellinger distance since it has the smallest computational cost in terms of number of operations.

The filtering procedure consists in checking which regions can be considered as coming from the same distribution that produced the data in the central block.
The sets which are not rejected are used to compute a local mean.
If all the sets are rejected, the filtered value is updated with the average on the central neighborhood around the filtered pixel.

\section{Results}\label{sec:assessment}

Image quality assessment in general, and filter performance evaluation in par\-tic\-u\-lar, are hard tasks~\cite{Moschetti2006}.
A ``good'' technique must combat noise and, at the same time, preserve details as well as relevant information.
In the following we assess the filters by two approaches.
Firstly, we use simulated data; with this, we are able to compare the true image (phantom) with the result of applying filters to corrupted version of the phantom.
Secondly, we apply measures of quality to a real image and its filtered version.

\subsection{Simulated data}\label{sec:SimulatedData}

\begin{figure}[hbt]
\centering
  \subfigure[Phantom]{\label{fig:phantom}%
  \includegraphics[width=.24\linewidth]{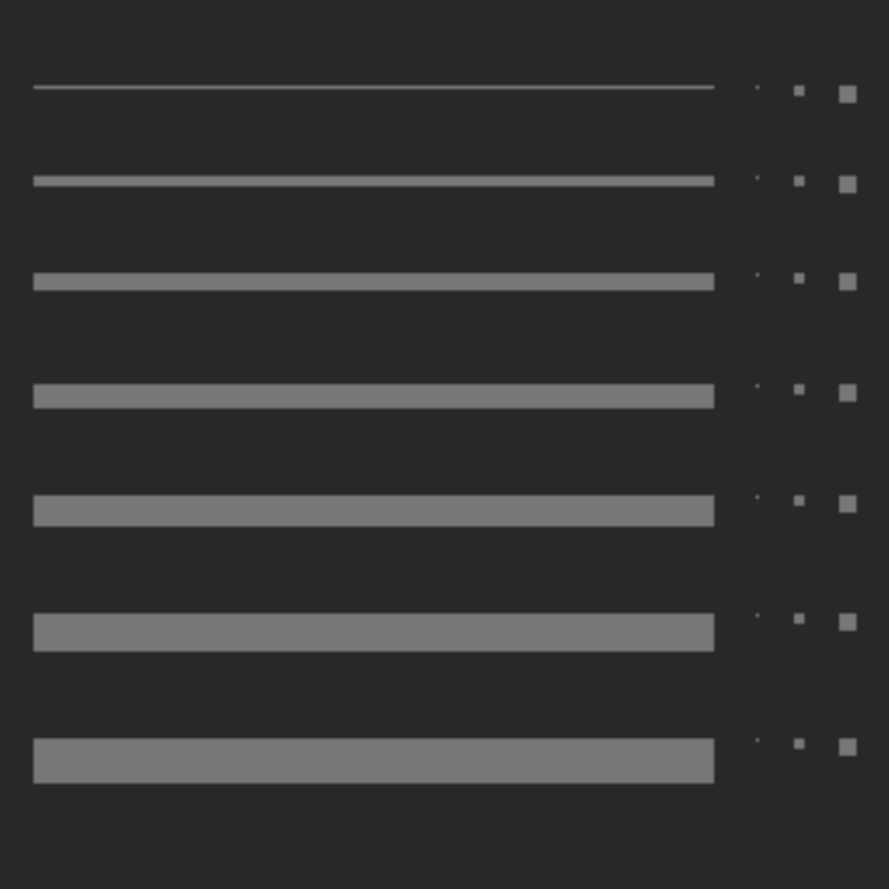}}%
  \subfigure[Corrupted]{\label{fig:corrupt}
  \includegraphics[width=.24\linewidth]{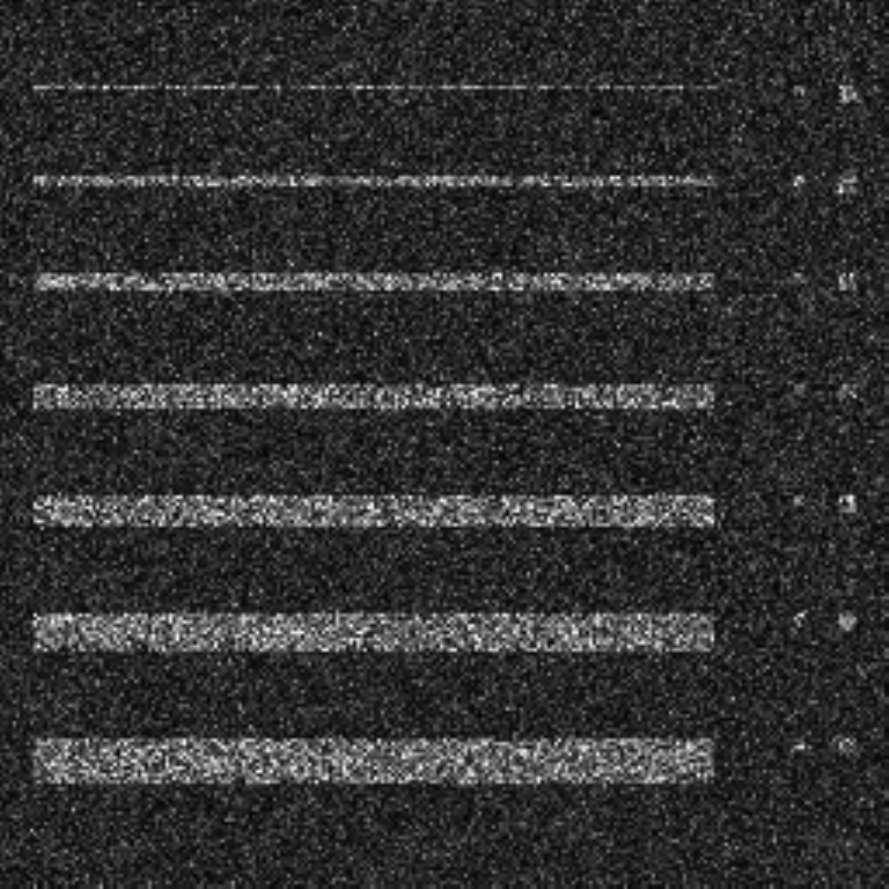}}%
  \subfigure[Lee's filter]{\label{fig:LeeFilter}
  \includegraphics[width=.24\linewidth]{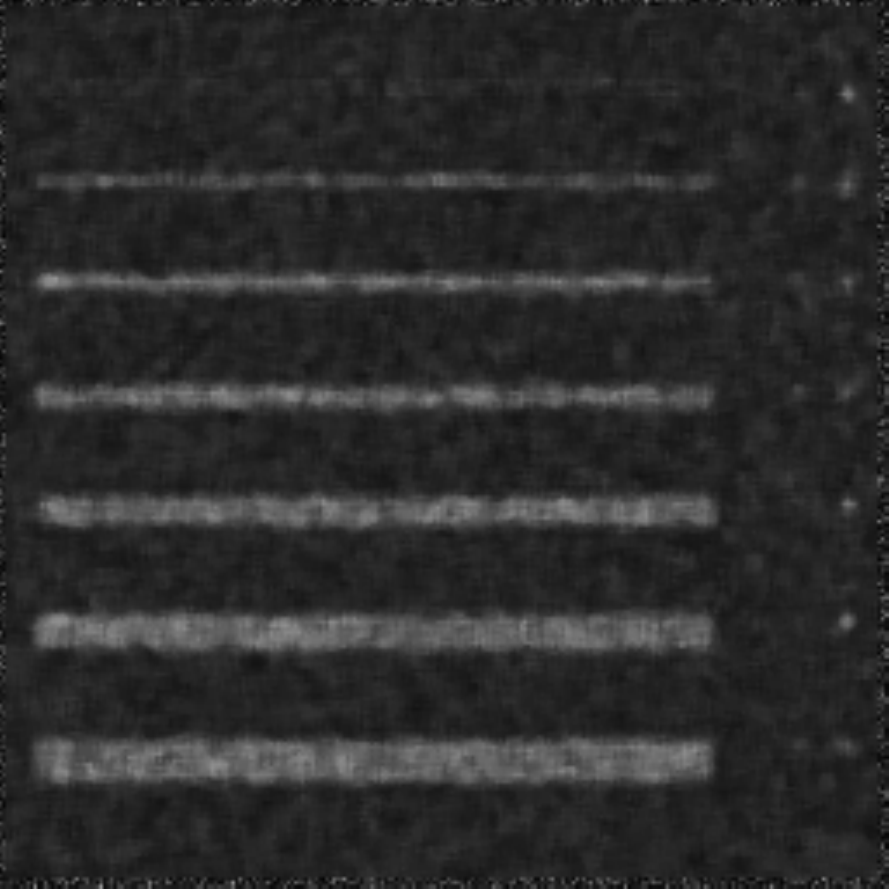}}%
  \subfigure[Hellinger filter]{\label{fig:HellingerFilter}
  \includegraphics[width=.24\linewidth]{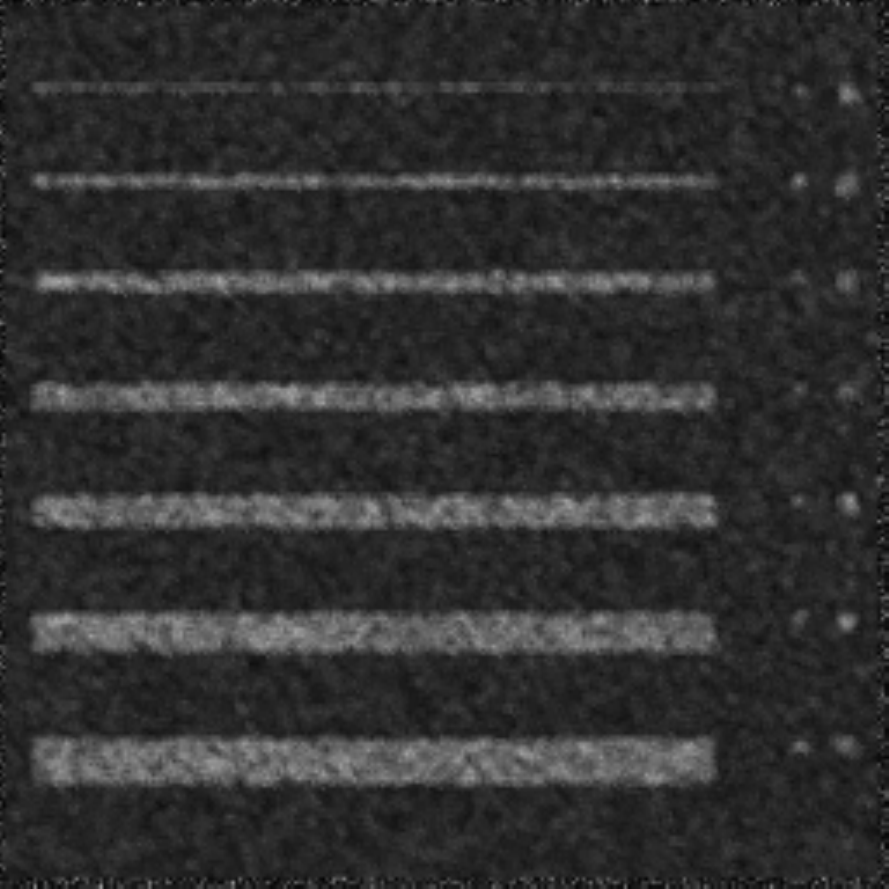}}%
\caption{Lee's Protocol, speckled data and filtered images.}
\label{fig:imgProtocol}
\end{figure}

The Monte Carlo experiment discussed in Moschetti et al.~\cite{Moschetti2006} consists of sim\-u\-lat\-ing corrupted images with different parameters.
Each simulated image is subjected to filters, and quality measures are computed from each result.
The quality of the filter with respect to each measure can then be assessed analyzing the data, not just a single image.
We use a phantom image (see Figure~\ref{fig:phantom}) which consists of light strips and points on a dark background, and we corrupt it with speckle noise (see Figure~\ref{fig:corrupt}).
The following measures of quality on the  filtered versions as, for instance, Figures~\ref{fig:LeeFilter} and~\ref{fig:HellingerFilter}, are then computed:
\begin{description}
 \item[Equivalent number of looks:] in intensity imagery and homogeneous areas, it can be estimated by $\textsf{NEL}=(\bar{z}/\widehat{\sigma}_Z)^2$, i.e., the square of the reciprocal of the coefficient of variation. In this case, the bigger the better.

 \item[Line contrast:] the pre\-serva\-tion of the line of one pixel width will be assessed by computing three means: in the coordinates of the original line ($x_{\ell}$) and in two lines around it ($x_{\ell_1}$ and $x_{\ell_2}$). The contrast is then defined as $2x_{\ell}-(x_{\ell_1} + x_{\ell_2})$, and compared with the contrast in the phantom. The best values are the smallest.

 \item[Edge preserving:] it is measured by means of the edge gradient (the absolute difference of the means of strip around edges) and variance (same as the former but using variances instead of means). The best values are the smallest.
 
 \item[The $Q$ index:] $Q = \frac{s_{xy}}{s_x s_y} \frac{2\overline{xy}}{\overline{x}^2 + \overline{y}^2} \frac{2 s_x s_y}{s_x^2 + s_y^2},$
where $s_\bullet^2$ and $\overline{\bullet}$ denote the sample variance and mean, respectively.
The range of $Q$ is $[-1,1]$, being $1$ the best value.

\item[The $\beta_{\rho}$ index:] $\beta_{\rho} = \frac{\sum_{j=1}^{n} (x_j-\bar{x})(y_j-\bar{y})}{\sqrt{\sum_{j=1}^{n} (x_j-\bar{x})^2 \sum_{j=1}^{n} (y_j-\bar{y})^2}},$
it is a correlation measure is between the Laplacians of images $X$ and $Y$, where $\bullet_j$ and $\overline{\bullet}$ denote the gra\-di\-ent values of the $jth$ pixel ​and mean ​of the images $\nabla^2X$ and $\nabla^2Y$, respectively.
The range of $\beta_{\rho}$ is $[-1,1]$, being $1$ the perfect correlation.

\end{description}

The proposed filter was compared with Lee's filter~\cite{Lee1986} which is considered a standard.
All filters were applied only once for the whole series of tests.
The results obtained are summarized by means of boxplots. 
Each boxplot describes the results of one filter applied to $100$ images formed by independent samples of the $\Gamma(L,L/\lambda)$ distribution with the parameters shown in Table~\ref{tab:simulated_data_I}.
These parameters describes situations usually found when analyzing SAR imagery in homogeneous regions.
The tests were performed at the $80\%$, $90\%$ and $99\%$ level of significance.

\begin{table}[hbt]
 \centering
 \renewcommand{\arraystretch}{.7} 
 \caption{Simulated situations with the $\Gamma(L,L/\lambda)$ distribution.}
 \begin{tabular}{c c c c} \toprule
  Situation ID & ~$L$~ & ~$\lambda$~ & Background mean \\ \midrule
       $\#1$     &  $1$  &  $200$      & $70$            \\ 
       $\#2$     &  $3$  &  $195$      & $55$            \\ 
       $\#3$     &  $5$  &  $150$      & $30$            \\ 
       $\#4$     &  $7$  &  $170$      & $35$            \\ \bottomrule
 \end{tabular}
 \label{tab:simulated_data_I}
\end{table}

\begin{figure}[hbt]
\centering
  \subfigure[Equivalent Number of Looks]{\label{fig:nel1}
  \includegraphics[width=.44\linewidth, trim=0 20 20 60]{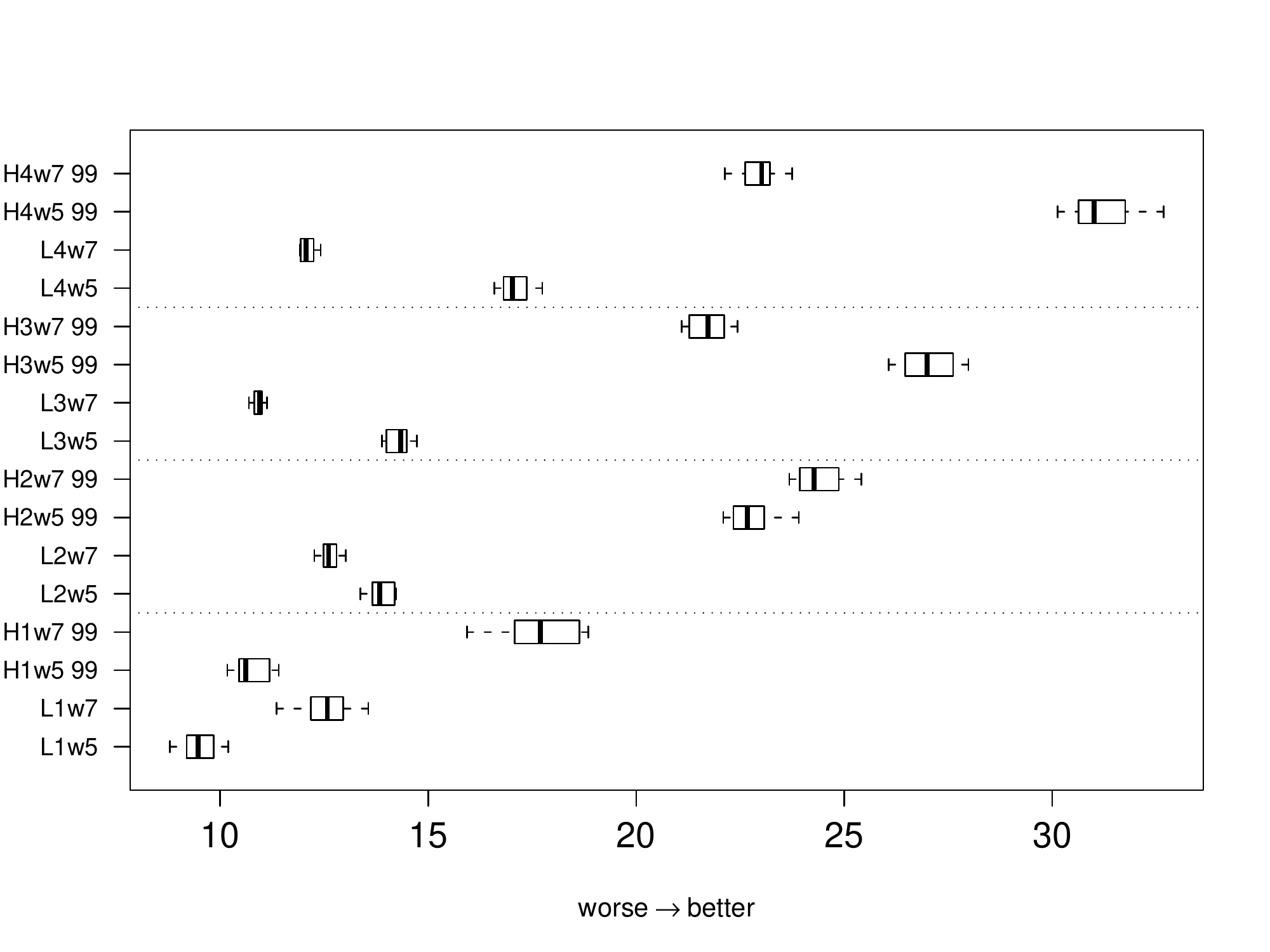}}
  \subfigure[Line Contrast]{\label{fig:line_pres1}
  \includegraphics[width=.44\linewidth, trim=0 20 20 60]{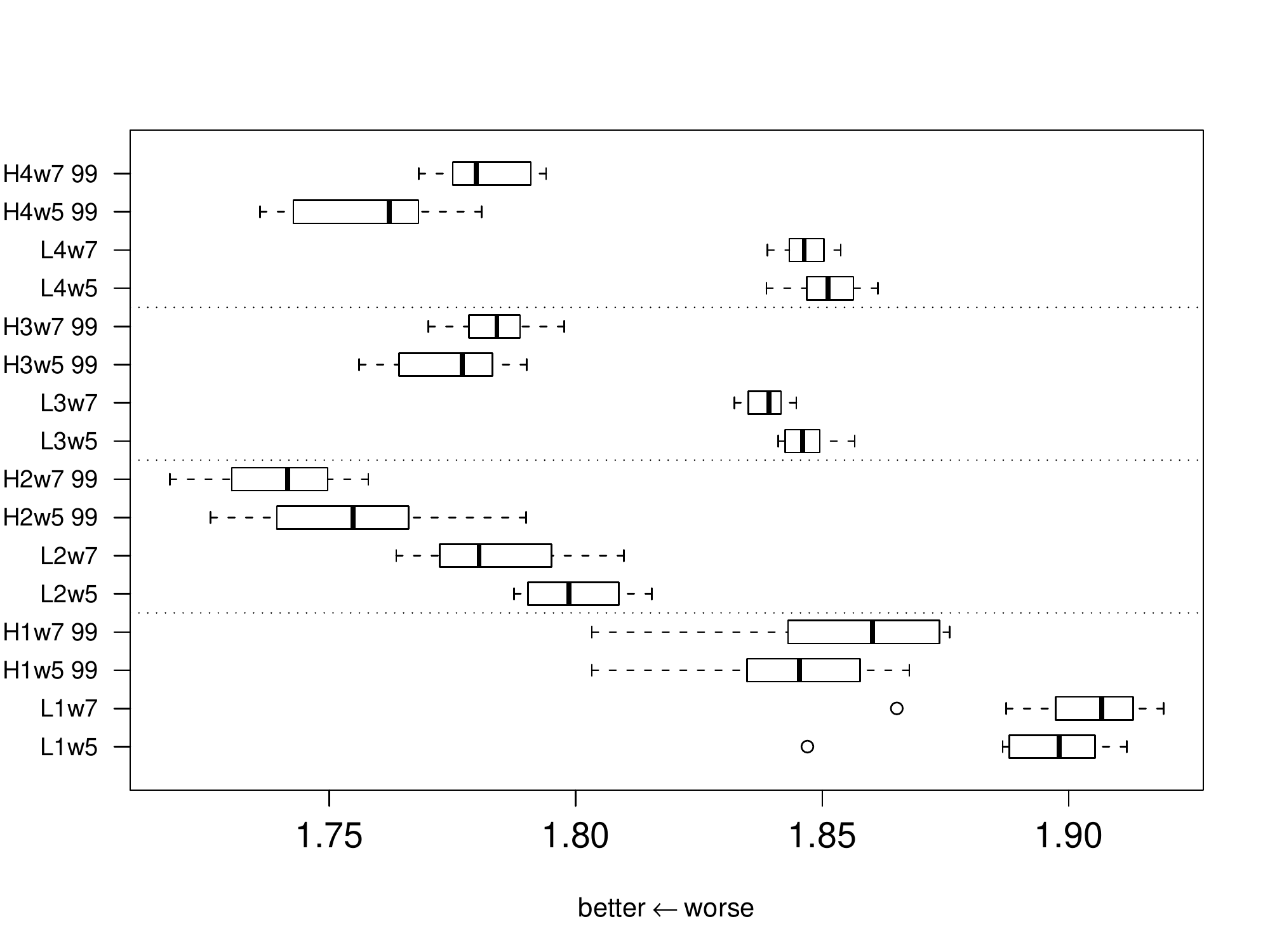}}\\
  
  \subfigure[Edge Gradient]{\label{fig:grad_mean1}
  \includegraphics[width=.44\linewidth, trim=0 20 20 60]{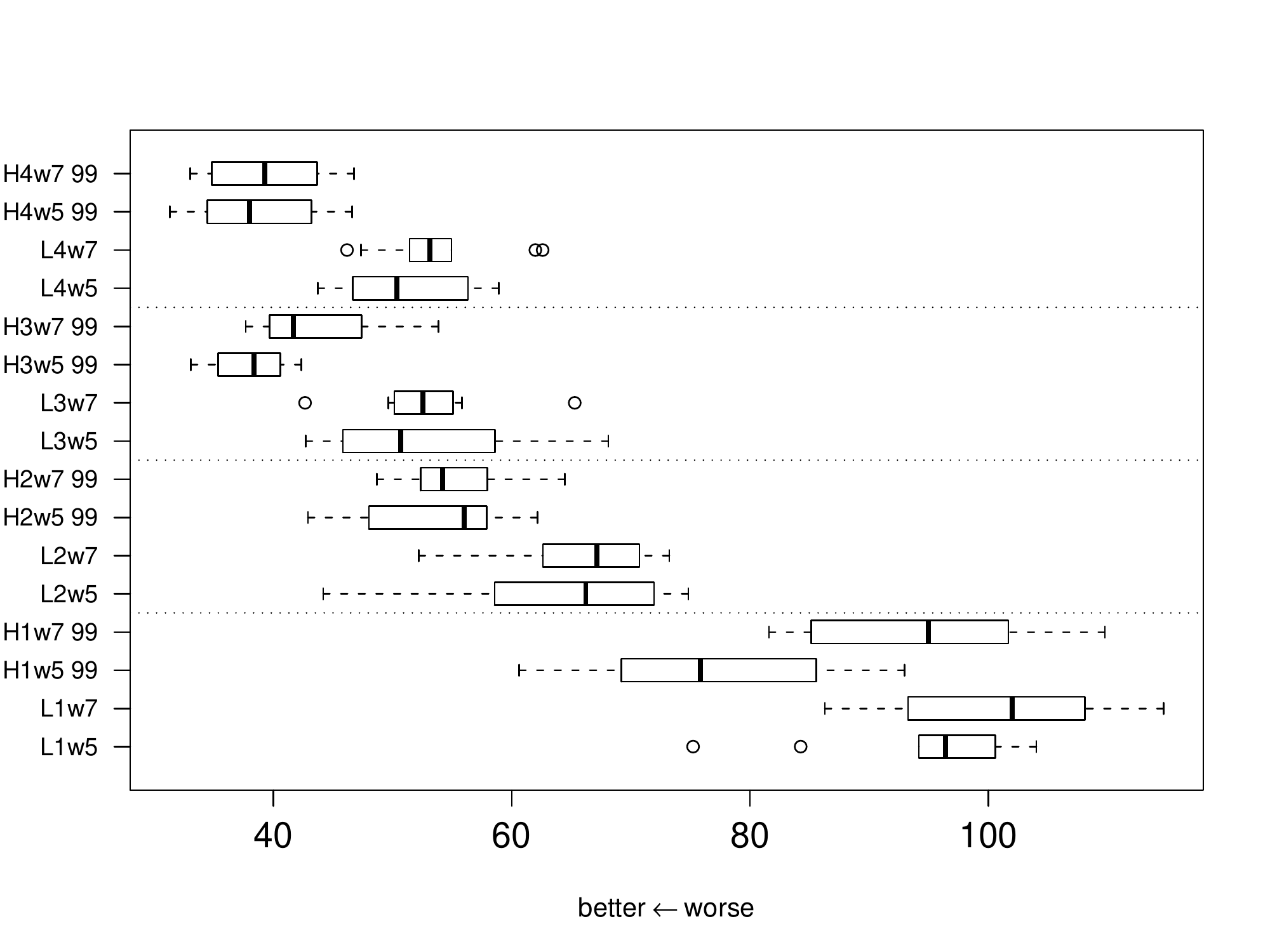}}
  \subfigure[Edge Variance]{\label{fig:grad_var1}
  \includegraphics[width=.44\linewidth, trim=0 20 20 60]{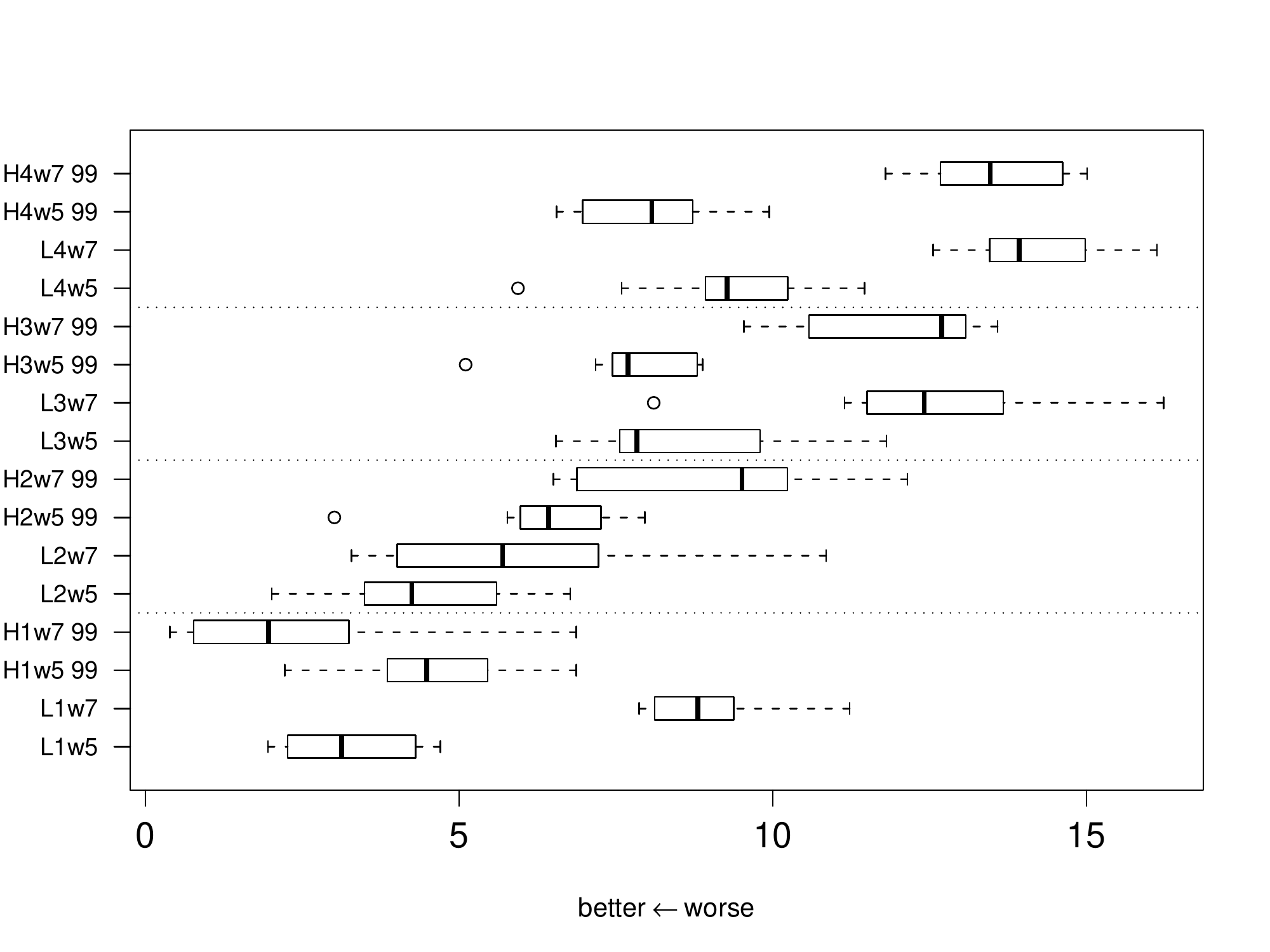}}\\

  \subfigure[Values of $Q$]{\label{fig:q_index1}
  \includegraphics[width=.44\linewidth, trim=0 20 20 60]{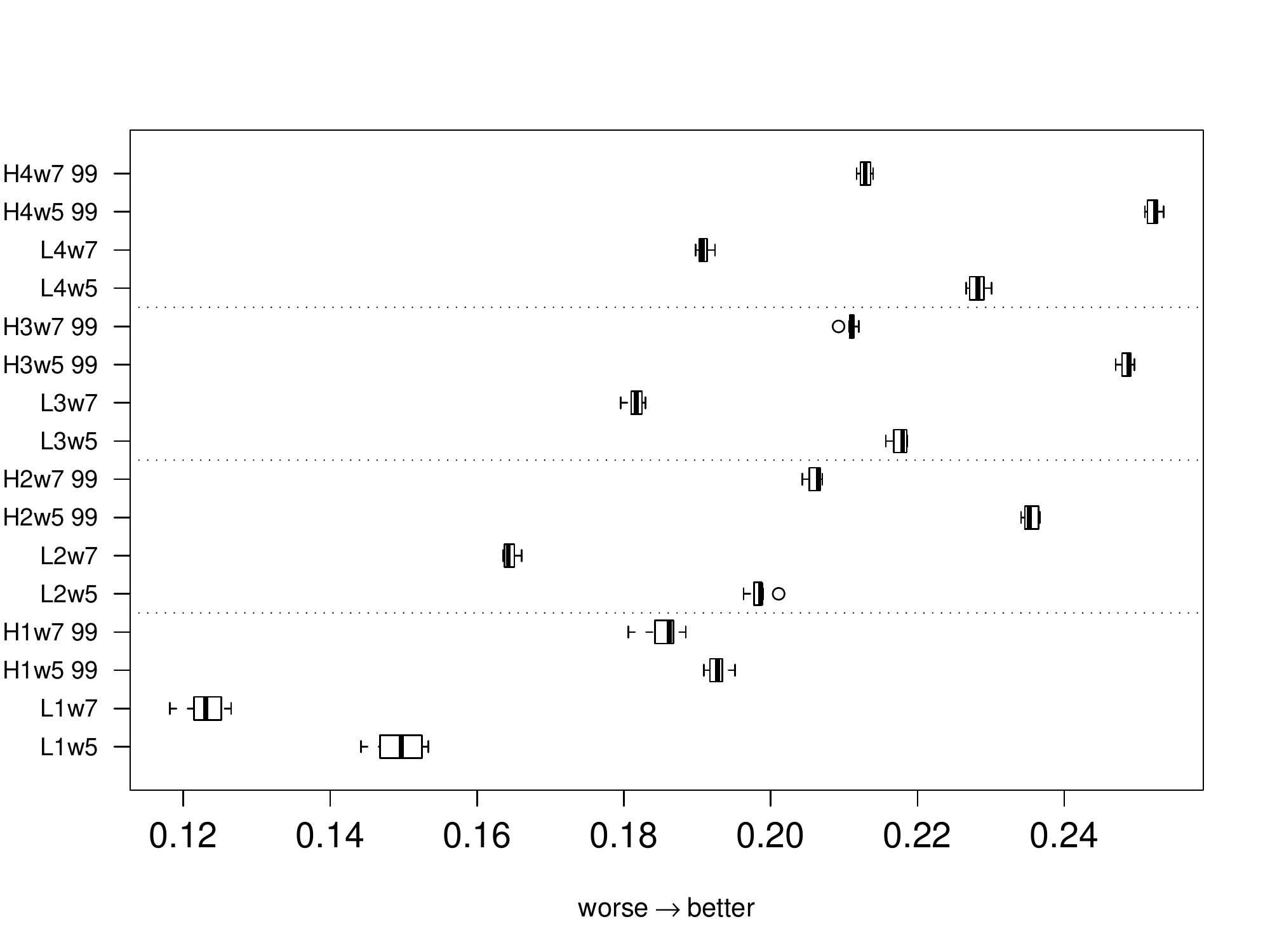}}
  \subfigure[Values of $\beta_{\rho}$]{\label{fig:beta_rho1}
  \includegraphics[width=.44\linewidth, trim=0 20 20 60]{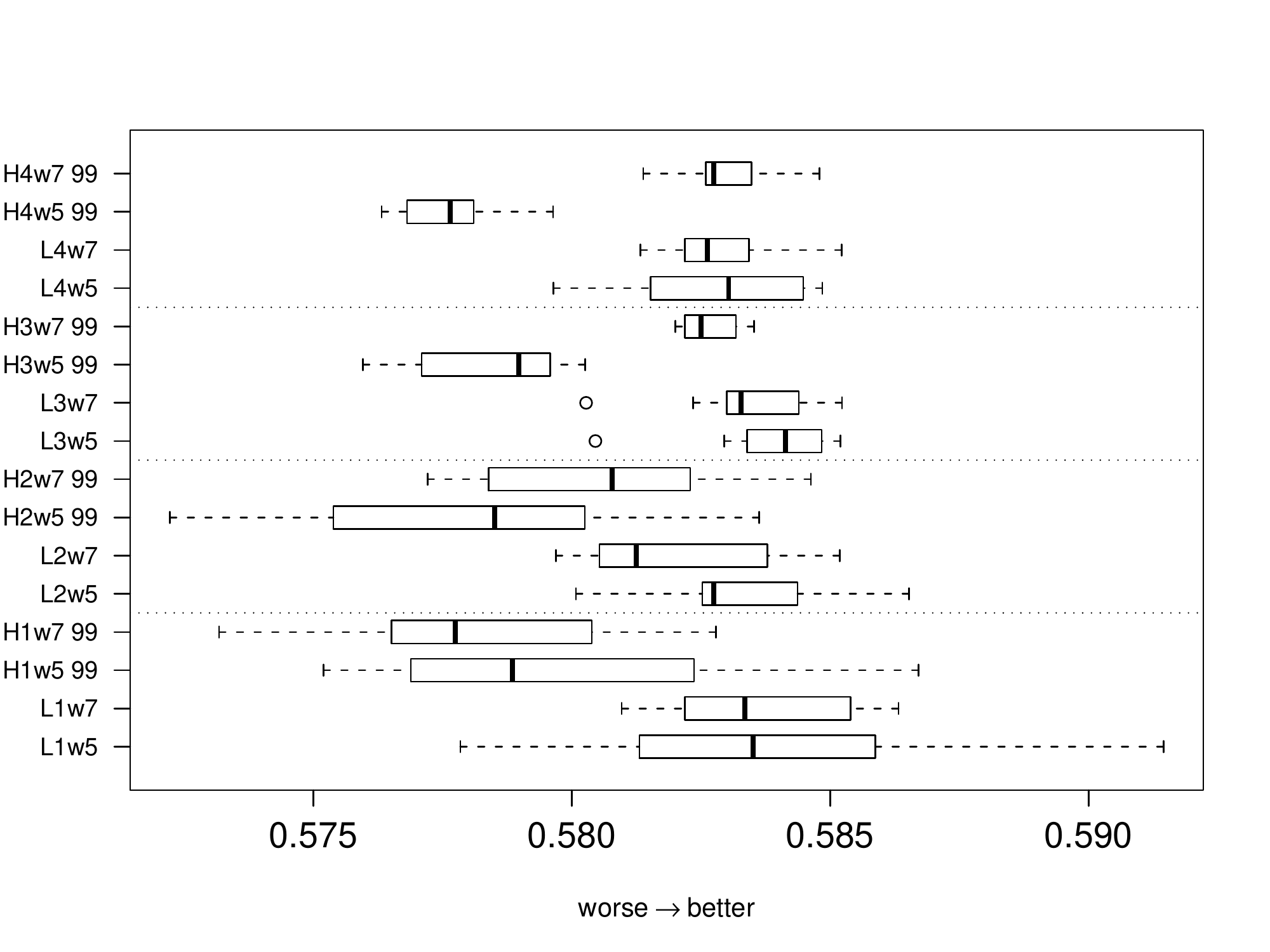}}
\caption{Boxplots of measures applied to four filters in four situations.}
\label{fig:boxplots_I}
\end{figure}

Figure~\ref{fig:boxplots_I} shows the boxplot of six measures corresponding to four filters.
Vertical axes are coded by the filter (`L' for Lee and `H' for Hellinger), the situation ID (from $1$ to $4$, see Table~\ref{tab:simulated_data_I}), the filter size ($5\times5$ and $7\times7$).
Only results at the $99\%$ level of significance are shown; the rest is consistent with this discussion.

Lee's filter presents better results than our proposal with respect to Edge Variance and the $\beta_\rho$ index in most of the considered situations, c.f.\ figures~\ref{fig:grad_var1} and~\ref{fig:beta_rho1}.
The filters based on stochastic distances consistently outperform Lee's filter with respect to all the other criteria, namely Number of Looks (figure~\ref{fig:nel1}), Line Contrast (figure~\ref{fig:line_pres1}), Edge Gradient (figure~\ref{fig:grad_mean1}), and the Universal Index Quality measure (figure~\ref{fig:q_index1}).

\subsection{Real data}\label{sec:RealData}

Not all the quality measures presented in Section~\ref{sec:SimulatedData} can be applied to real data, unless the ground truth is known.
For this reason, the following metrics will be used in this case~\cite{SARNumericalMeasures}, where the smaller is the better (they are all error measures):
\begin{description}

 \item[Mean Absolute Error:] $\textsf{MAE} = {n}^{-1}\sum_{j=1}^{n} | x_j-y_j |$.

 \item[Mean Square Error:] $\textsf{MSE} = {n}^{-1}\sum_{j=1}^{n} (x_j-y_j)^2$.

 \item[Normalized Mean Square Error:] $\textsf{NMSE} = \frac{\sum_{j=1}^{n} (x_j-y_j)^2}{\sum_{j=1}^{n} x_j^2}$. 

 \item[Distortion Contrast:] $\textsf{DCON} = {n}^{-1}\sum_{j=1}^{n} \frac{\arrowvert x_j-y_j \arrowvert}{\alpha+x_i+y_j}$, where $\alpha$ depends on the relationship between luminance and gray level of the display; we used $\alpha = 23/255$.
\end{description}

\begin{figure}[hbt]
\centering
  \subfigure[SAR data]{\label{fig:SARdata}
  \includegraphics[width=.24\linewidth]{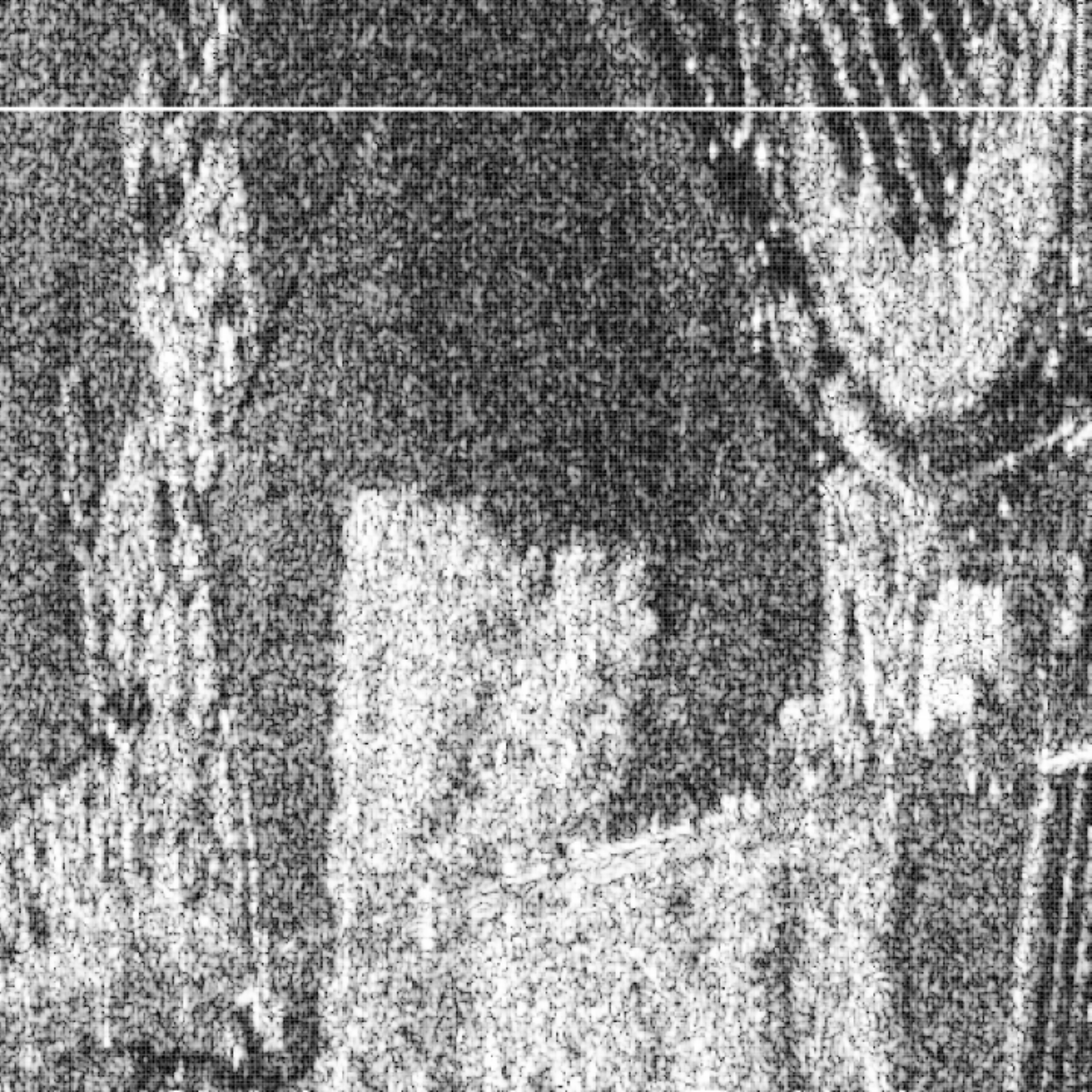}}%
  \subfigure[Lee filter]{\label{fig:SAR_LeeFilter}
  \includegraphics[width=.24\linewidth]{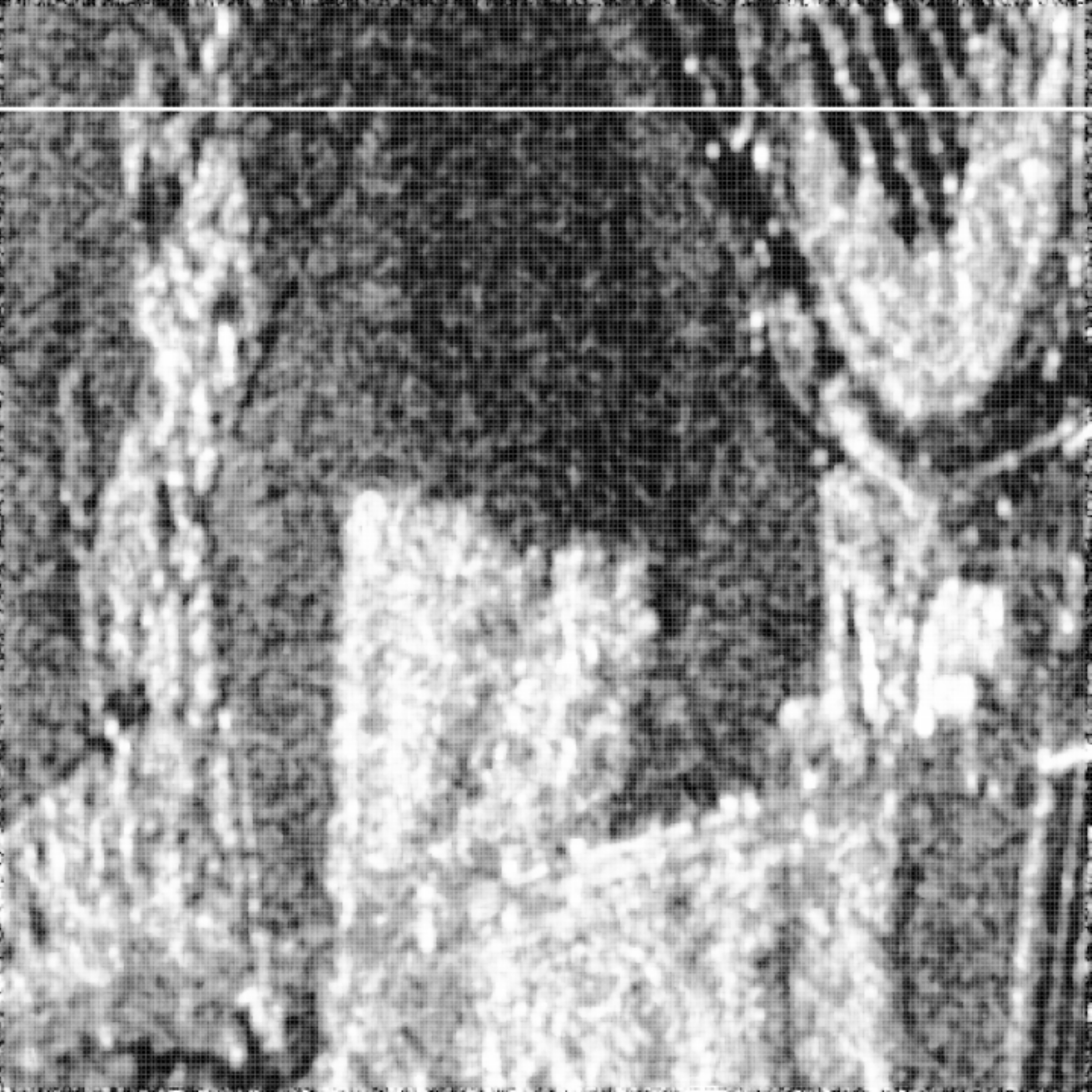}}%
  \subfigure[Hellinger filter]{\label{fig:SAR_HellingerFilterW5}
  \includegraphics[width=.24\linewidth]{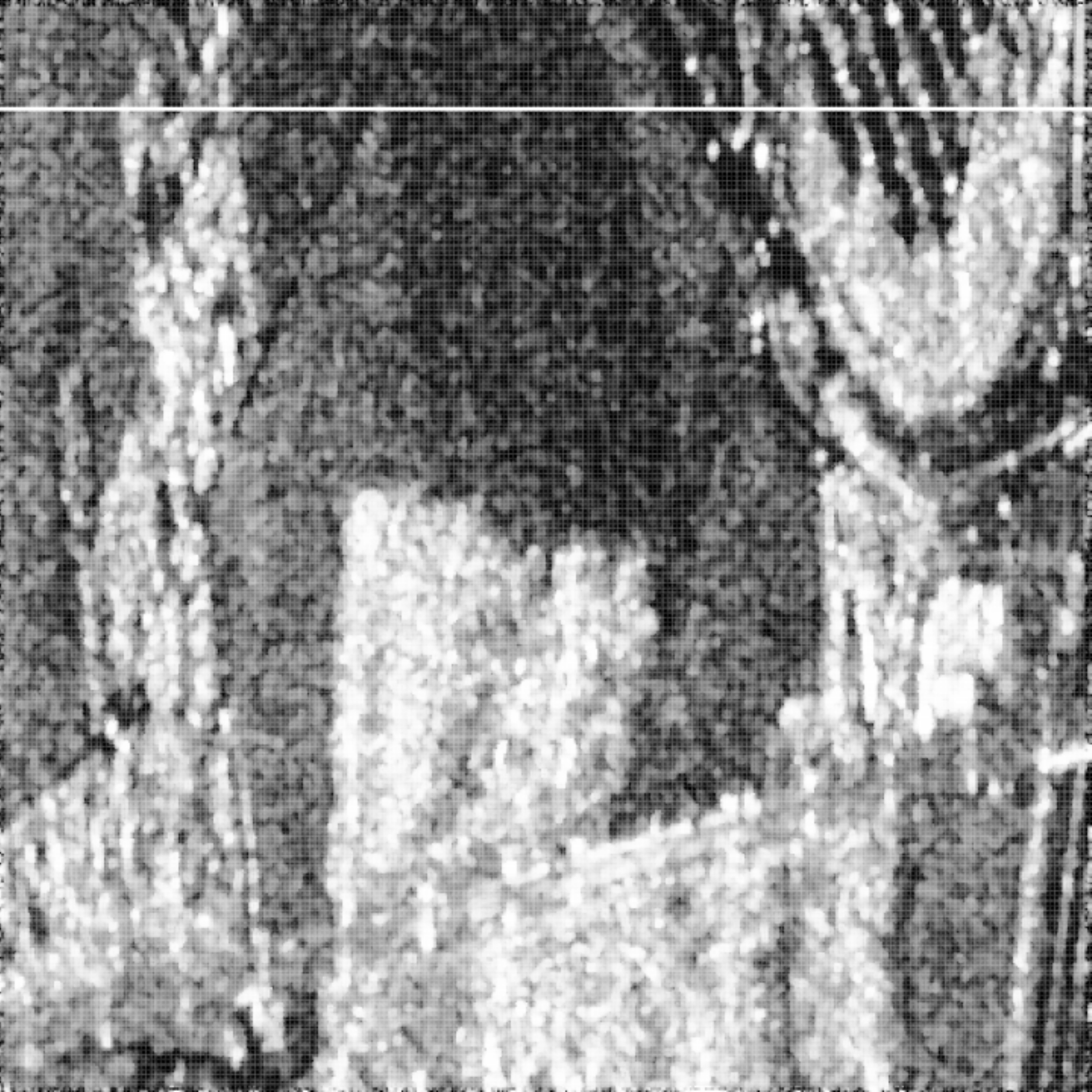}}%
  \subfigure[Profile]{\label{fig:analysis_1D}
  \includegraphics[width=.27\linewidth]{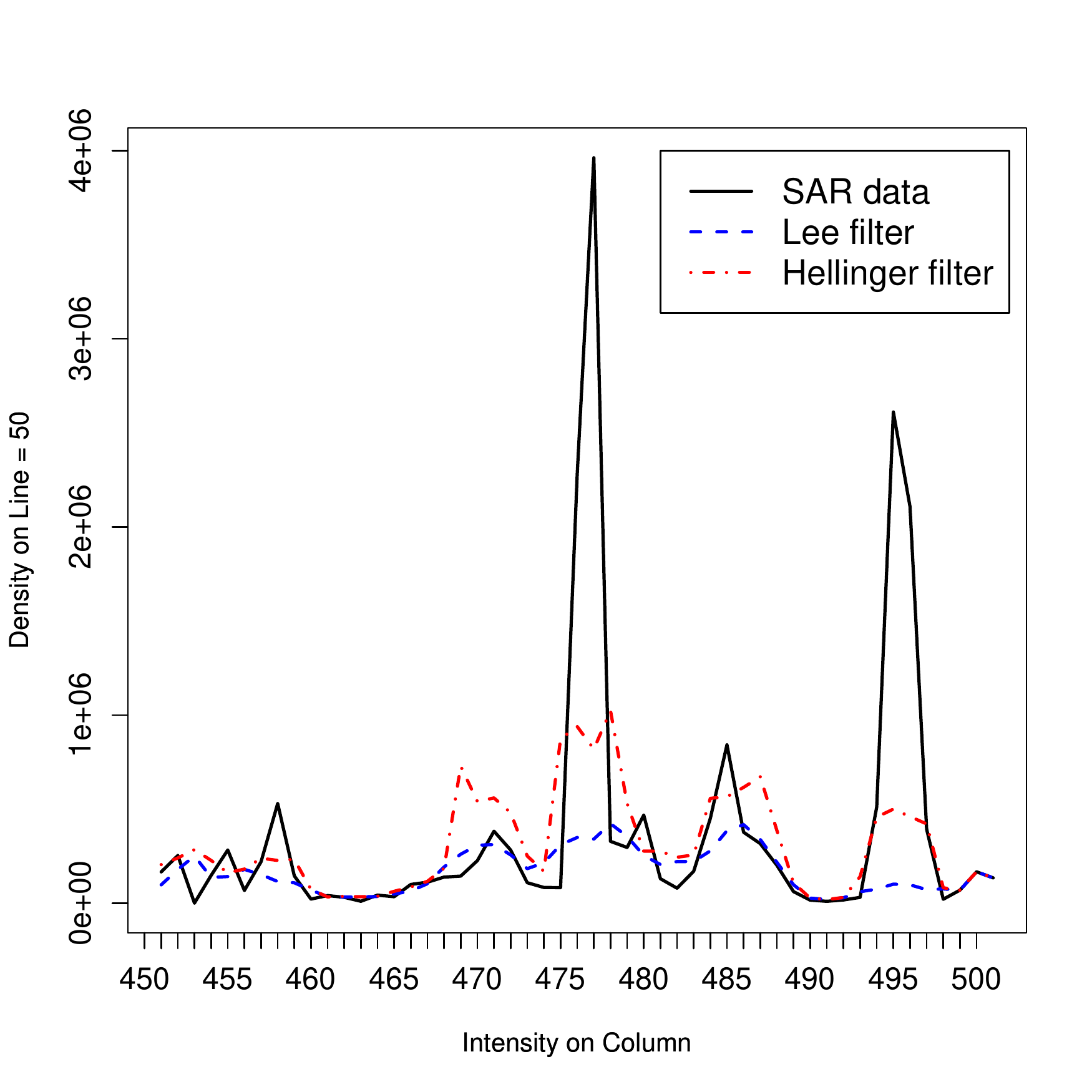}}%
\caption{SAR data, filtered images and $1$D analysis.}
\label{fig:SARdataSituations}
\end{figure}

Figure~\ref{fig:SARdataSituations} presents the real image, its filtered versions and analysis $1$-D of the $row=50$.
The original data were produced by the E-SAR sensor in the L band (HH polarization) with \unit[$2.2\times 3.0$]{m} of ground resolution and four nominal looks.
Nascimento et al.~\cite{Nascimento2010} analyzed this image, and the equivalent number of looks in homogeneous areas is always below three.

The Lee filtered image is smoother that the ones obtained with stochastic distances, but comparing figures~\ref{fig:SAR_LeeFilter} and~\ref{fig:SAR_HellingerFilterW5} one notices that our proposal retains much more detail than the classical technique.
Figure~\ref{fig:analysis_1D} presents the profile of the images in the highlighted line.
While combating speckle with almost the same success: the bright spots in the upper right corner, which are seldom visible after applying the Lee filter, stand out in the image filtered with the Hellinger distance and windows of side $5$ at a level $\alpha=80\%$.

Table~\ref{tab:SARdataMeasures} presents the assessment of the filters, and we note that the Hellinger filter of order $5$ with $\alpha=80\%$ achieved the best results.

\begin{table}[hbt]
 \centering
 \renewcommand{\arraystretch}{.7}
 \caption{Image quality indexes in the real SAR image.}
 \begin{tabular}{c c c c c c c c} 
  \toprule
    & \textbf{Speckle} & \multicolumn{4}{c}{Measures of Quality} & \multicolumn{2}{c}{$Q$ index} \\ \cmidrule(lr{.5em}){3-6} \cmidrule(lr{.5em}){7-8} 
$\bm{\alpha}$ & \textbf{Filter}  & \multirow{1}{*}{\textsf{~MAE~}} & \multirow{1}{*}{\textsf{~MSE~}} & \multirow{1}{*}{\textsf{NMSE~}} & \multirow{1}{*}{\textsf{DCON}} & \multirow{1}{*}{$\bar{Q}$} & \multirow{1}{*}{~~~$s_{Q}$~~~} \\ 
   \midrule
   &Lee\_w5& 0.145 & 0.037 & 0.110 & 0.184 & 0.142 & 0.138 \\ \cmidrule(lr{.5em}){2-8}
   &Lee\_w7& 0.156 & 0.042 & 0.126 & 0.195 & 0.082 & 0.127 \\ \midrule
   \multirow{2}{*}{$\bm{80\%}$} 
   & H\_w5 & \textbf{0.117} & \textbf{0.025} & \textbf{0.076} & \textbf{0.155} & \textbf{0.486} & 0.170 \\ \cmidrule(lr{.5em}){2-8} 
   & H\_w7 & 0.141 & 0.035 & 0.104 & 0.180 & 0.265 & 0.187 \\ \midrule
   \multirow{2}{*}{$\bm{90\%}$} 
   & H\_w5 & 0.120 & 0.026 & 0.080 & 0.159 & 0.453 & 0.176 \\ \cmidrule(lr{.5em}){2-8} 
   & H\_w7 & 0.142 & 0.035 & 0.106 & 0.182 & 0.250 & 0.189 \\ \midrule
   \multirow{2}{*}{$\bm{99\%}$} 
   & H\_w5 & 0.127 & 0.029 & 0.085 & 0.166 & 0.397 & 0.180 \\ \cmidrule(lr{.5em}){2-8} 
   & H\_w7 & 0.145 & 0.036 & 0.109 & 0.185 & 0.222 & 0.189 \\ 
  \bottomrule
 \end{tabular} 
 \label{tab:SARdataMeasures}
\end{table}

\section{Conclusions}\label{sec:conclu}

We presented new filters based on stochastic distances for speckle noise reduction.
The proposal was compared with the classical Lee filter, using a protocol based on Monte Carlo experiences, showing that it is competitive. 
An applications to real SAR data was presented and, numerical methods were used to assert the proposal.
The proposed filters behave nearly alike, and they outperform the Lee filter in almost all quality measures.
However, other significance levels will be tested, along with different points of parameter space to have a yet more complete assessment of proposal.
The proposal can be extended to any problem, requiring only the computation of stochastic distances.

\bibliographystyle{splncs03}
\bibliography{ref_TorresCIARP2012,ref_imageQualityIndex}

\begin{thebibliography}{10}
\providecommand{\url}[1]{\texttt{#1}}
\providecommand{\urlprefix}{URL }

\bibitem{EstimationEquivalentNumberLooksSAR}
Anfinsen, S.N., Doulgeris, A.P., Eltoft, T.: Estimation of the equivalent
  number of looks in polarimetric synthetic aperture radar imagery. IEEE
  Transactions on Geoscience and Remote Sensing  47(11),  3795--3809 (2009)

\bibitem{SARNumericalMeasures}
Baxter, R., Seibert, M.: Synthetic aperture radar image coding. {MIT} Lincoln
  Laboratory Journal  11(2),  121--158 (1998)

\bibitem{Gao2010}
Gao, G.: Statistical modeling of {SAR} images: {A Survey}. Sensors  10,
  775--795 (Jan 2010)

\bibitem{Goodman1976}
Goodman, J.W.: Some fundamental properties of speckle. Journal of the Optical
  Society of America  66(11),  1145--1150 (1976)

\bibitem{Lee1986}
Lee, J.S.: Speckle suppression and analysis for synthetic aperture radar
  images. Optical Engineering  25(5),  636--645 (1986), iSSN 0091-3286

\bibitem{Moschetti2006}
Moschetti, E., Palacio, M.G., Picco, M., Bustos, O.H., Frery, A.C.: On the use
  of {L}ee's protocol for speckle-reducing techniques. Latin American Applied
  Research  36(2),  115--121 (Apr 2006)

\bibitem{NagaoMatsuyama}
Nagao, M., Matsuyama, T.: Edge preserving smoothing. Computer Graphics and
  Image Processing  9(4),  394--407 (Apr 1979)

\bibitem{Nascimento2010}
Nascimento, A.D.C., Cintra, R.J., Frery, A.C.: Hypothesis testing in speckled
  data with stochastic distances. IEEE Transactions on Geoscience and Remote
  Sensing  48(1),  373--385 (Jan 2010)

\bibitem{Salicru1994}
Salicr{\'u}, M., Morales, D., Men{\'e}ndez, M.L., Pardo, L.: On the
  applications of divergence type measures in testing statistical hypotheses.
  Journal of Multivariate Analysis  21(2),  372--391 (Nov 1994)

\bibitem{UIQIndex}
Wang, Z., Bovik, A.C.: A universal image quality index. {IEEE} Signal Process.
  Letters  9(3),  81--84 (2002)

\end{thebibliography}

\end{document}